\documentclass[preprint]{elsarticle}

\begin{document}

\begin{frontmatter}

\title{Future Colliders for Particle Physics --- ``Big and Small''}
\author{Frank Zimmermann\thanks{frank.zimmermann@cern.ch}, CERN, Geneva, Switzerland}


\begin{abstract}
Discoveries at high-energy particle colliders have established the standard model of 
particle physics.
Technological innovation has helped to increase the collider energy at a much 
faster pace than the corresponding costs.
New concepts will allow reaching ever higher luminosities and energies throughout the coming century.
Cost-effective strategies for the collider implementation include staging.
For example, a future circular collider could first provide electron-positron collisions, 
then hadron collisions (proton-proton and heavy-ion), and finally the collision of muons.
Cooling-free muon colliders, realizable in a number of ways, promise an attractive
and energy-efficient path towards  lepton collisions at tens of TeV.
While plasma accelerators and dielectric accelerators offer unprecedented gradients,
the construction of a high-energy collider based on these new 
technologies still calls for significant improvements in cost and performance.     
Pushing the accelerating gradients or bending fields 
ever further, the breakdown of the QED vacuum may set an ultimate limit to  
electromagnetic acceleration. 
Finally, some ideas are sketched for reaching, or exceeding, the Planck energy.
\end{abstract}


\end{frontmatter}

\section{Introduction}
In January 1932 the first practical 
cyclotron of E.O.~Lawrence with a diameter of 32 cm  
accelerated protons to a kinetic energy of 1.22 
MeV. The beam current was about 1 nA \cite{Livingston}.
Almost 90 years later the Large Hadron Collider (LHC) at 
CERN, with a diameter of 9 km, accelerates
protons to an energy of 6.5 TeV 
with a beam current of the order of 1 A. 
Therefore, over about a century, the accelerators became $3\times 10^{4}$ larger
and now achieve close to $10^{7}$ times higher energy than 100 years ago. 
While a large part of this energy rise was accomplished by an  increase in size,  
two to three orders of magnitude were gained by technological improvements. 

At the same time, the technology advances have also dramatically reduced
the specific construction cost, i.e. the cost per centre-of-mass (c.m.) energy, 
from about 60 MCHF/(GeV c.m.) for the CERN Proton Synchotron (PS),
first operated in 1959,  
to 0.3 MCHF/(GeV c.m.) for the Large Hadron Collider (LHC), 
in operation since 2008, with both prices expressed in 
2008 Swiss francs [CHF] \cite{lebrun,Zimmermann:1708761} --- 
a cost reduction by a factor 200 over 40 years.  

\begin{figure}[htb]
\centering
\includegraphics[width=0.8\columnwidth]{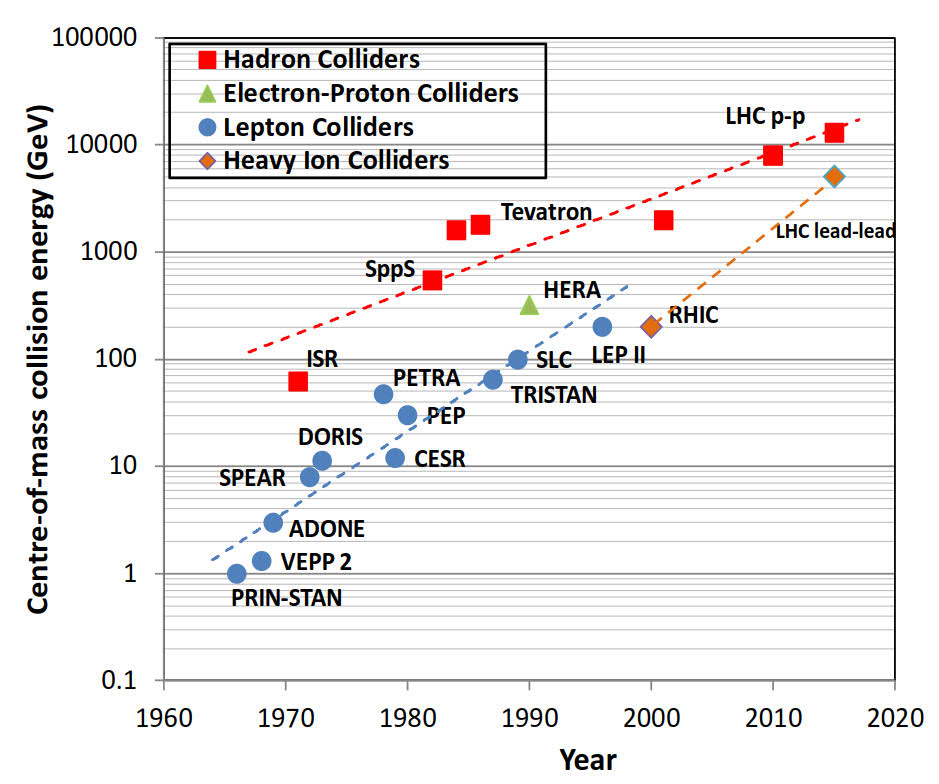}
\caption{Centre-of-mass energy of particle colliders versus year 
\protect\cite{shiltsev,ballarino}.} 
\label{fig1}
\end{figure}

The colliders, which could be realized thanks to this advancing technology, have proven 
powerful instruments for discovery and precision measurement. All the heavier particles of the standard model were produced 
at colliders: 
the tau lepton and charm quark at SPEAR, the top quark at the Tevatron,
the gluon at PETRA, the W and Z bosons at the S$p\bar{p}$S collider,
and the Higgs boson at the LHC.
These and other colliders appear in Fig.~\ref{fig1}.

The past progress was enabled by the introduction of new concepts (e.g., strong focusing, separate function magnets, colliding beams) as much as  
by the emergence of new technologies, 
in particular ones based on superconductivity.
PETRA, TRISTAN and LEP-II started the massive use of superconducting 
radiofrequency (rf) systems.
The Tevatron was the first accelerator based on superconducting magnets.
HERA, RHIC, and the LHC used, or use, both superconducting magnets and superconducting rf.

In addition to energy and specific cost, also the 
accelerator performance was tremendously improved over time:  
Every year the LHC delivers more luminosity 
than all the previous hadron colliders together
had accumulated over their entire operating history.

\section{Pushing the Energy Frontier in the 21st Century}
A very large circular hadron collider appears to be the 
only feasible approach to reach 100 TeV c.m. collision energy in the 
coming decades. Such collider would offer 
access to new particles through direct production  
in the few-TeV to 30 TeV mass range, far beyond the LHC reach \cite{Mangano:2270978}.
It would also provide 
much-increased rates 
for phenomena in the sub-TeV mass range and,
thereby, a much increased precision compared with the LHC \cite{Mangano:2270978}.  

The centre-of-mass 
energy reach of a hadron collider is directly proportional to the maximum magnetic field $B$ 
and to the bending radius $\rho$:  
\begin{equation}
E_{\rm c.m.} \propto \rho  B .
\end{equation}
Therefore, an increase in the size of the collider compared with the 
LHC by a factor of about 4 and an approximate doubling of the magnetic 
field yields almost an order of magnitude increase in energy.

Such approach was first suggested in the year 2010  
during the High-Energy LHC (HE-LHC) workshop \cite{Todesco:1344820}. 
Now it is the focus of the Future Circular Collider (FCC) study \cite{benedikt}, 
which was launched in response to the 2013 Update of the 
European Strategy for Particle Physics:
16 Tesla magnets in a 100 km ring will result in a centre-of-mass energy of 100 TeV. 
This goal defines the overall infrastructure requirements for the FCC accelerator complex.
The FCC study scope also includes the design of a high-luminosity 
e$^+$e$^-$ collider (FCC-ee), as a possible first step  --- with a remarkably rich 
physics programme \cite{tlepphysics} ---,  
as well as a proton-electron collision option (FCC-he) at one interaction point, 
where a 60 GeV electron beam from an energy recovery linac 
is collided with one of the two 50 TeV proton beams circulating in the FCC-hh.
The design of a higher-energy hadron collider in the LHC tunnel 
based on FCC-hh magnet technology 
--- the so-called High-Energy LHC (HE-LHC) 
--- is yet another part of the FCC study.

As of June 2017, 111 institutes and 25 companies from 30 countries are participating in the FCC study effort.
The near-term goal is to deliver a conceptual design report of 
all FCC collider options, including technologies, detector design, and physics goals,
before the end of 2018, as input to the next European
Strategy Update process expected for 2019/20.

\begin{figure}[htbp]
\centering
\includegraphics[width=0.9\columnwidth]{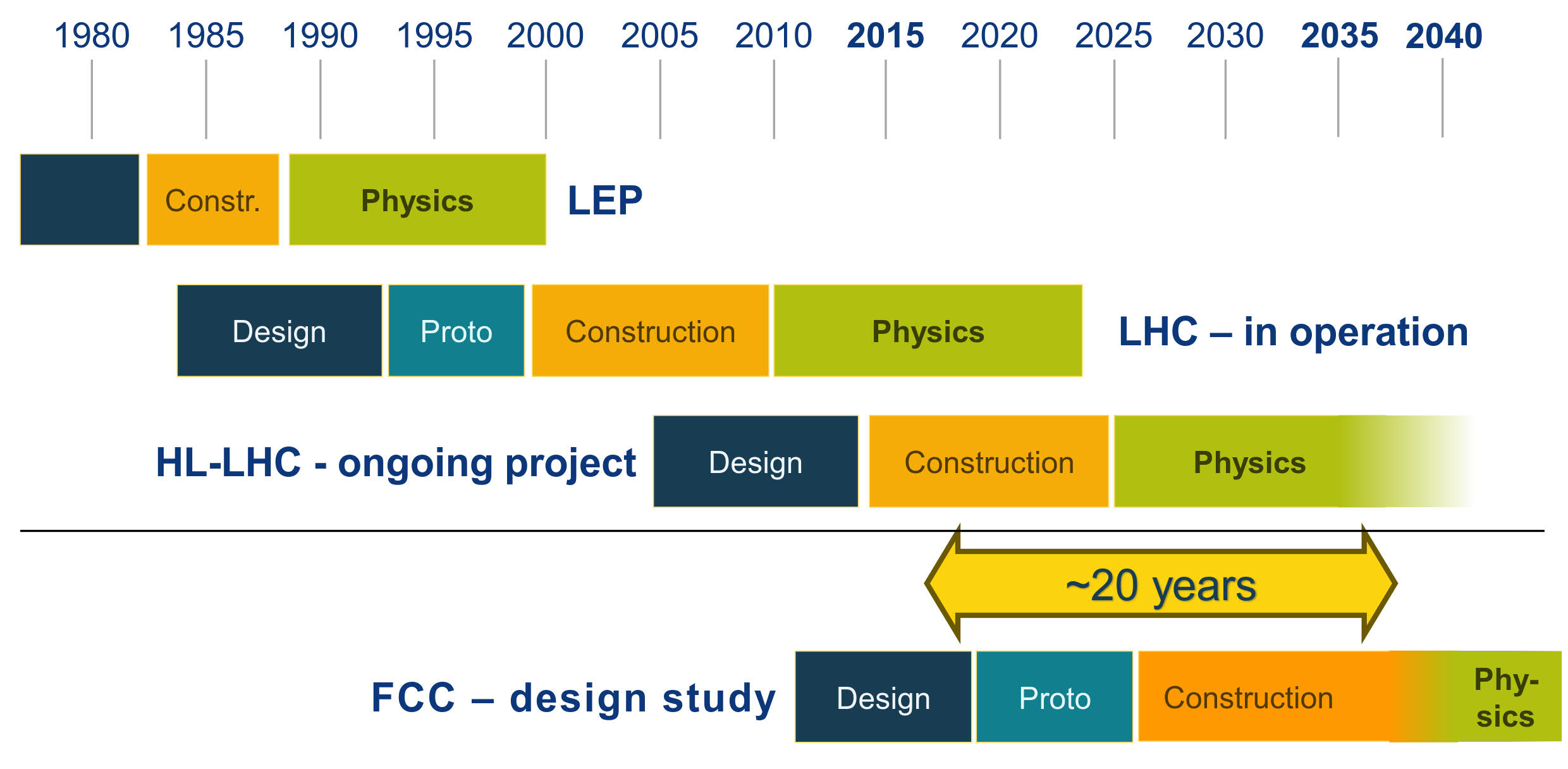}
\caption{Time lines of several 
past, present and future circular colliders at CERN, distinguishing periods of design, prototyping, construction, and physics exploitation.}
\label{fig2}
\end{figure}

Figure \ref{fig2} compares the time lines of various past 
and present circular colliders at CERN with a projected time line for the FCC, indicating a need for fast progress.

CEPC and SppC are two colliders similar to FCC-ee/FCC-hh,
which are being studied by another international collaboration, centred at IHEP Beijing \cite{accel:cepc}. 
These two machines have a similar circumference of about 100 km. 
Several possible locations in China are under study.  
The e$^+$e$^-$ collider CEPC is designed with a maximum centre-of-mass energy of 240 GeV, and a 
3 or 200 times lower luminosity than FCC-ee, at the Higgs production peak and on the Z resonance, respectively.  
The SppC hadron collider relies on 12 T (later 24 T) iron-based high-temperature superconducting magnets, 
which could be installed in the same tunnel as the CEPC.  

Table \ref{tabhadron} shows key parameters of FCC-hh, SppC, and HE-LHC, together  
with the design values of the present LHC and its luminosity upgrade (HL-LHC).
Table \ref{tablepton} compares parameters for FCC-ee and CEPC at different operating
energies with those of LEP-2 and SuperKEKB.

\begin{table}[htbp]
\begin{center}
\begin{tabular}{|l|c|c|c|c|c|c|}
\hline
parameter & 
\multicolumn{2}{|c|}{FCC-hh} & 
\multicolumn{2}{|c|}{SppC} & 
HE-LHC
& (HL-)LHC \\
\hline 
c.m.~energy [TeV] 
& \multicolumn{2}{|c|}{100} 
& 75 & 150 
& 27 & 14 \\
dipole field  [T] 
& \multicolumn{2}{|c|}{16} 
& 12 & 24 
& 16 & 8.3 \\
\hline 
circumference [km] 
  & \multicolumn{2}{|c|}{97.8} 
& \multicolumn{2}{|c|}{100} 
& 26.7 & 26.7 \\
\hline 
beam current [A] 
  & \multicolumn{2}{|c|}{0.5} 
& 0.77 & --  
& 1.12 & (1.12) 0.58  \\
part./bunch [$10^{11}$] 
  & \multicolumn{2}{|c|}{1} 
& 1.5 & --  
& 2.2 & (2.2) 1.15  \\
bunch spacing [ns]  
  & \multicolumn{2}{|c|}{25} 
& 25 & -- 
& 25 & 25 \\
norm.~emittance [$\mu$m]  
  & \multicolumn{2}{|c|}{2.2 (1.1)}
& 3.16 & -- 
& 2.5 (1.25) & (2.5) 3.75 \\
\hline 
IP beta function [m]  
  & 1.1 & 0.3 
& 0.71 & --  
& 0.25 & (0.15) 0.55 \\
lum.~[$10^{34}$ cm$^{-2}$s$^{-1}$]  
  & 5 & 30  
& 10 & 100  
& 25 & (5) 1 \\
events per crossing
  & 170 & 1000  
& $\sim$300 &  --  
& 800 & (135) 27 \\
\hline 
SR power/beam [kW]
& \multicolumn{2}{|c|}{2400} 
& 1130 & --  
& 100 & (7.3) 3.6 \\
longit.~damp.~time [h]
& \multicolumn{2}{|c|}{1.1} 
& 2.4 &  --  
& 3.6 & 25.8 \\
\hline
init.~burn-off time [h]
& 17 & 3.4
& 13 & --   
& 3.0 & (15) 40 \\
\hline
\end{tabular}
\end{center}
\caption{Parameters of future hadron colliders, 
the LHC and its HL-LHC upgrade}
\label{tabhadron} 
\end{table}

\begin{table}[htbp]
\begin{center}
\begin{tabular}{|l|c|c|c|c|c|c|c|}
\hline
parameter & 
\multicolumn{3}{|c|}{FCC-ee} & 
\multicolumn{2}{|c|}{CEPC} & 
LEP-2
& SKEKB \\
\hline
beam energy [GeV] 
& 45.6 & 120 & 182.5 & 45.5 & 120 & 104 & 
4/7 \\
\hline 
circumference [km] 
  & \multicolumn{3}{|c|}{97.8} 
& \multicolumn{2}{|c|}{100} 
& 26.7 & 3.0  \\
\hline 
beam current [mA] 
  & 1390 & 29 & 5.4 & 
19 & 74 & 
3 & $\sim$3000 \\
part./bunch [$10^{11}$] 
 & 1.7 & 1.5 & 2.8 
&  0.1 & 1.0  
& 4.2 &  $\sim 1$ 
\\
h.~emittance  [nm]  
  & 0.3  & 0.6  & 2 & 
0.2 & 1.3  
& 22 & 4 
\\
v.~emittance  [pm]  
  & 1  & 1  & 3 & 
 0.9 & 4   
& 250 & 10 
\\
h.~IP beta [m]  
  & 0.15 &  0.3 & 1 &  
0.17 &  0.17 &
1.2  & 0.03 \\
v.~IP beta [mm]  
  & 0.8 &  1 & 2 &  
2 &  2 &
50  & 0.3 \\
lum.~[$10^{34}$ cm$^{-2}$s$^{-1}$]    
  & $>$200 & $>$7 & $>$1.3    
& 1.1 & 2.0  
& 0.01  & 80 \\
\hline
\end{tabular}
\end{center}
\caption{Parameters of future e$^+$e$^-$ colliders, 
LEP-2 and SuperKEKB.}
\label{tablepton} 
\end{table}

Construction cost of
 future projects can, and should, be 
minimized by \cite{shiltsev2,zimmermann2}:  
(1) reducing the cost of the components,
e.g., superconductor and high-field magnet for the
hadron colliders;
(2) building on a site with an existing infrastructure
and injector complex, e.g., on the CERN site;
and 
(3) staging, e.g., FCC-ee followed by FCC-hh, and possibly
by a muon collider FCC-$\mu\mu$ (see below).

\section{Mitigating Synchrotron Radiation}
Synchrotron radiation (SR) is an obstacle on the path to higher 
energy, as the resulting energy loss per turn 
increases with the fourth power of a charged particle's energy. 
For FCC-ee, synchrotron radiation 
limits the maximum attainable c.m.~energy to
about 400 GeV since at higher energy the required 
rf voltage could no longer be provided by conventional
rf technology.   
Also, by design, 
the FCC-ee beams emit a total of 100 MW
SR power at all energies.

However, synchrotron radiation does not
only afflict the electron and positron machines.
The synchrotron radiation of the FCC-hh hadron collider
amounts to about 5 MW, emitted inside the 
cryogenic arcs. 
The power consumption of the FCC-hh hadron collider
is likely to be dominated by the effects of this synchrotron radiation. 
Namely, the heat extraction from the beam screen  
inside the cold magnets and from the cold bore itself 
requires well above 100 MW of electric power
for the cryogenics plants.
In addition, of order 10 MW electric 
power is required for the rf system 
to maintain the energy of the stored beams.

Several possible approaches exist for eliminating, avoiding, 
or at least reducing the synchrotron radiation:
\begin{itemize} 
\item 
suppression of synchrotron radiation for circular 
e$^{+}$e$^{-}$ and hadron colliders by (1) 
shaping the beam:  
classically a time-invariant beam --- like a constant-current loop ---
does not radiate, i.e., it does not emit any electromagnetic waves;  
for the same reason a quieter beam 
or a crystalline beam would radiate less  \cite{ratner,primack};  
and/or by (2) tayloring the beam-pipe boundary: 
a large bending radius $\rho$ combined with a small chamber 
suppresses SR emission 
at long wavelengths; specifically, radiation is shielded at 
wavelengths $\lambda\ge 2\sqrt{d^3/\rho}$, 
where $d$ denotes the beam-pipe diameter \cite{warnock};
\item
transiting from circular to linear e$^{+}$e$^{-}$ colliders, such as 
the proposed International Linear Collider (ILC) 
\cite{ilc} or Compact Linear Collider (CLIC) \cite{clic}; 
however, linear colliders
are intrinsically inefficient, as the entire beam is disposed,
and its energy lost, after only a single collision;
on the other hand, synchrotron radiation is still present during the 
collision (here called ``beamstrahlung''), where it becomes 
a significant limitation at multi-TeV beam energies;
future advanced variants of linear colliders with higher gradient, 
and therefore of smaller size, are proposed to be based on 
laser- or beam-driven plasma wake-field acceleration
schemes (LPA \cite{lpa} or PWFA \cite{pwfa,pwfa1,pwfa2}) or on 
dielectric laser 
acceleration (DLA \cite{dla}) schemes; 
at present such acceleration schemes, which promise  
10--1000 times higher accelerating gradients than conventional rf structures,  
still tend to have a non-negligible specific cost, currently  
as high as the super- or normalconducting rf systems \cite{shiltsev2}, 
if not higher; their  
luminosity per wall-plug power 
might also increase with beam energy less strongly   
than required by particle physics \cite{delahaye}; 
preserving the quality of a positron beam during plasma
acceleration may prove yet another challenge \cite{pwfap}; 
\item
changing the particle type from e$^{+}$e$^{-}$ to muons;
indeed such muon-collider schemes \cite{delahaye} promise a high efficiency, as 
we will discuss in the following.
\end{itemize}

\section{Maximizing Efficiency}
\begin{figure}[htbp]
\centering
\includegraphics[width=0.9\columnwidth]{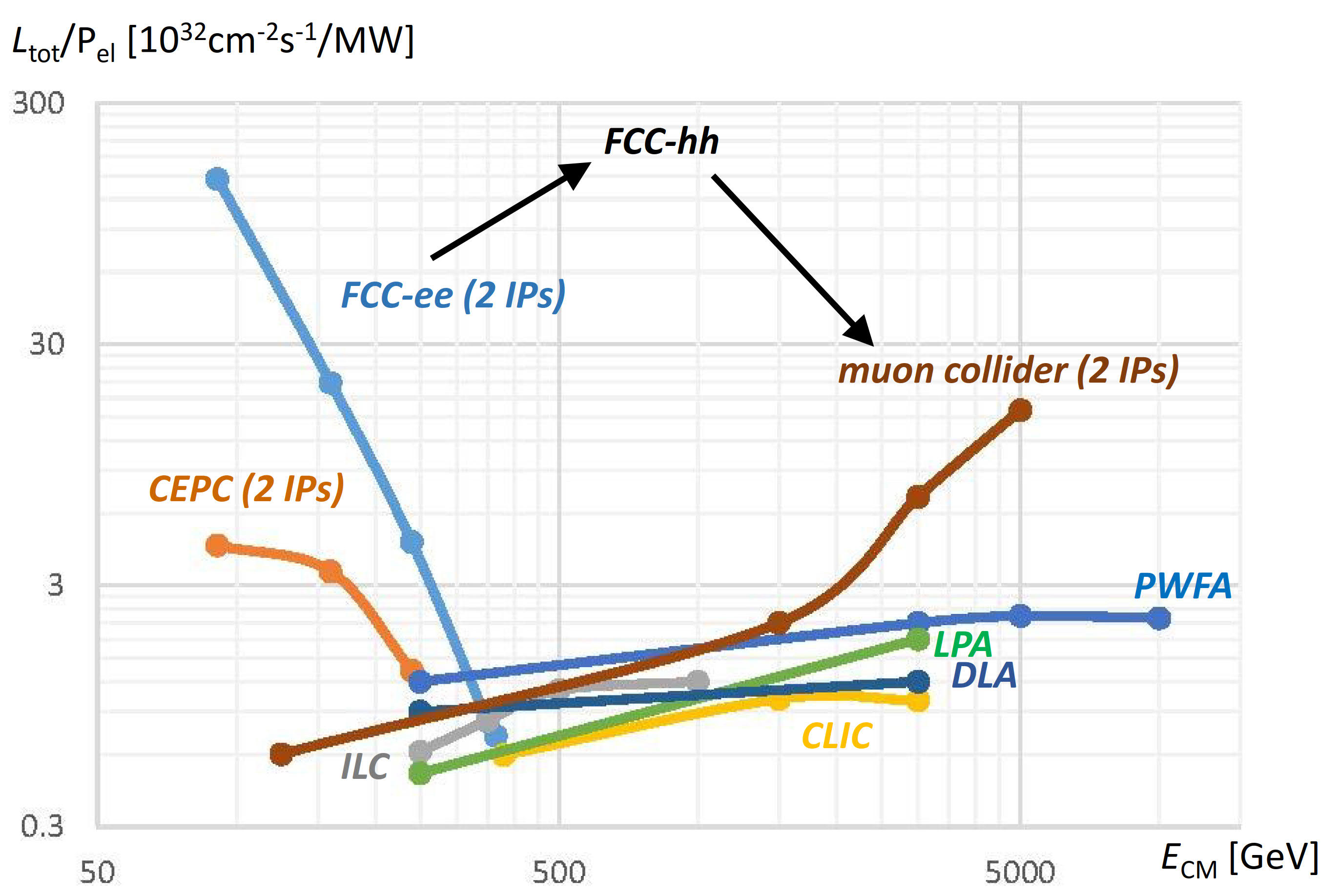}
\caption{Lepton-collider 
luminosity per electrical power for different accelerator projects and technologies,
hinting that the staged approach FCC-ee ($\rightarrow$ FCC-hh) $\rightarrow$ 
FCC-$\mu\mu$ would deliver highest possible 
luminosity at c.m.~energies from 91 GeV to $>5$  TeV in the most efficient way 
\cite{delahaye,abpj}.  
}
\label{fig3}
\end{figure}

Figure \ref{fig3} 
demonstrates that up to the top quark threshold the circular collider FCC-ee offers 
by far the highest luminosity at the lowest electric input power,
revealing the FCC-ee characteristics as a truly ``green'' accelerator.
The FCC-ee would naturally be followed by FCC-hh, which  
spans a wide range of parton collision energies 
at high luminosity, for a total electric input power similar to the one of FCC-ee.
A possible third stage would be a muon collider, the only 
lepton-collider approach which promises a high  
luminosity at higher energy \cite{delahaye}, 
as is required for particle-physics explorations.

\begin{figure}[htbp]
\centering
\includegraphics[width=0.90\columnwidth]{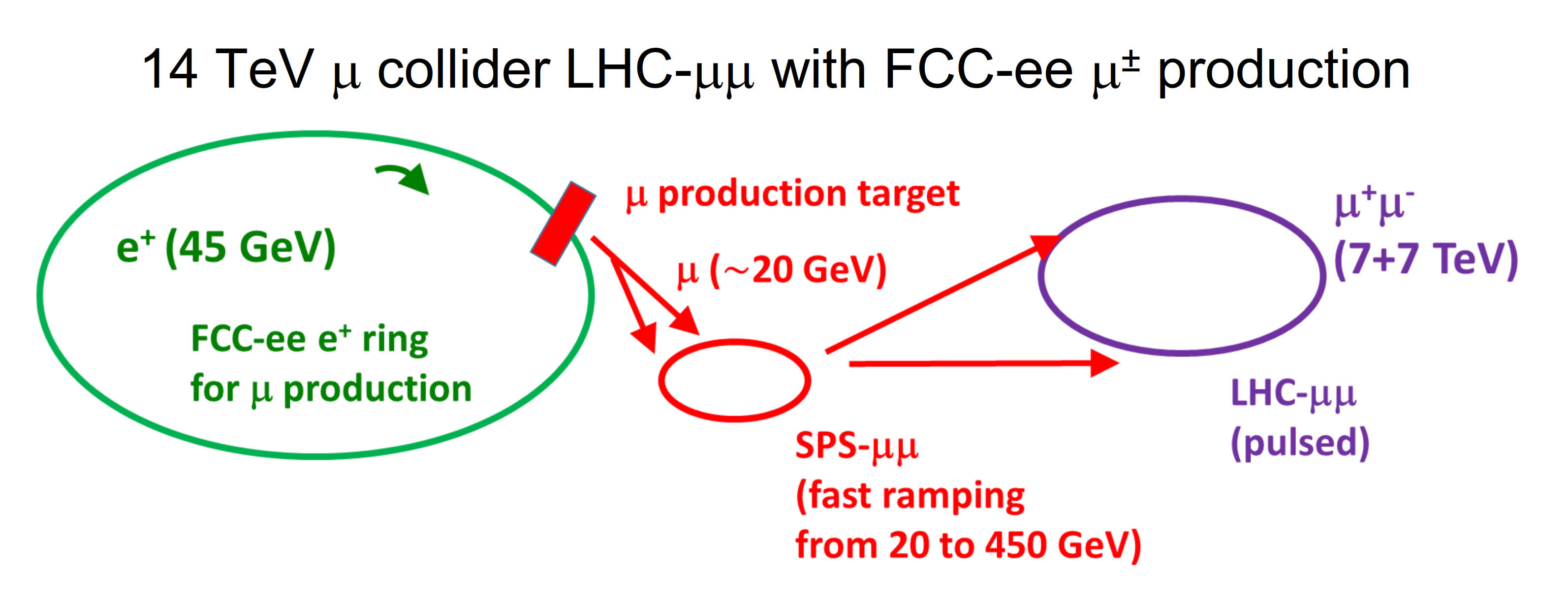}
\includegraphics[width=0.90\columnwidth]{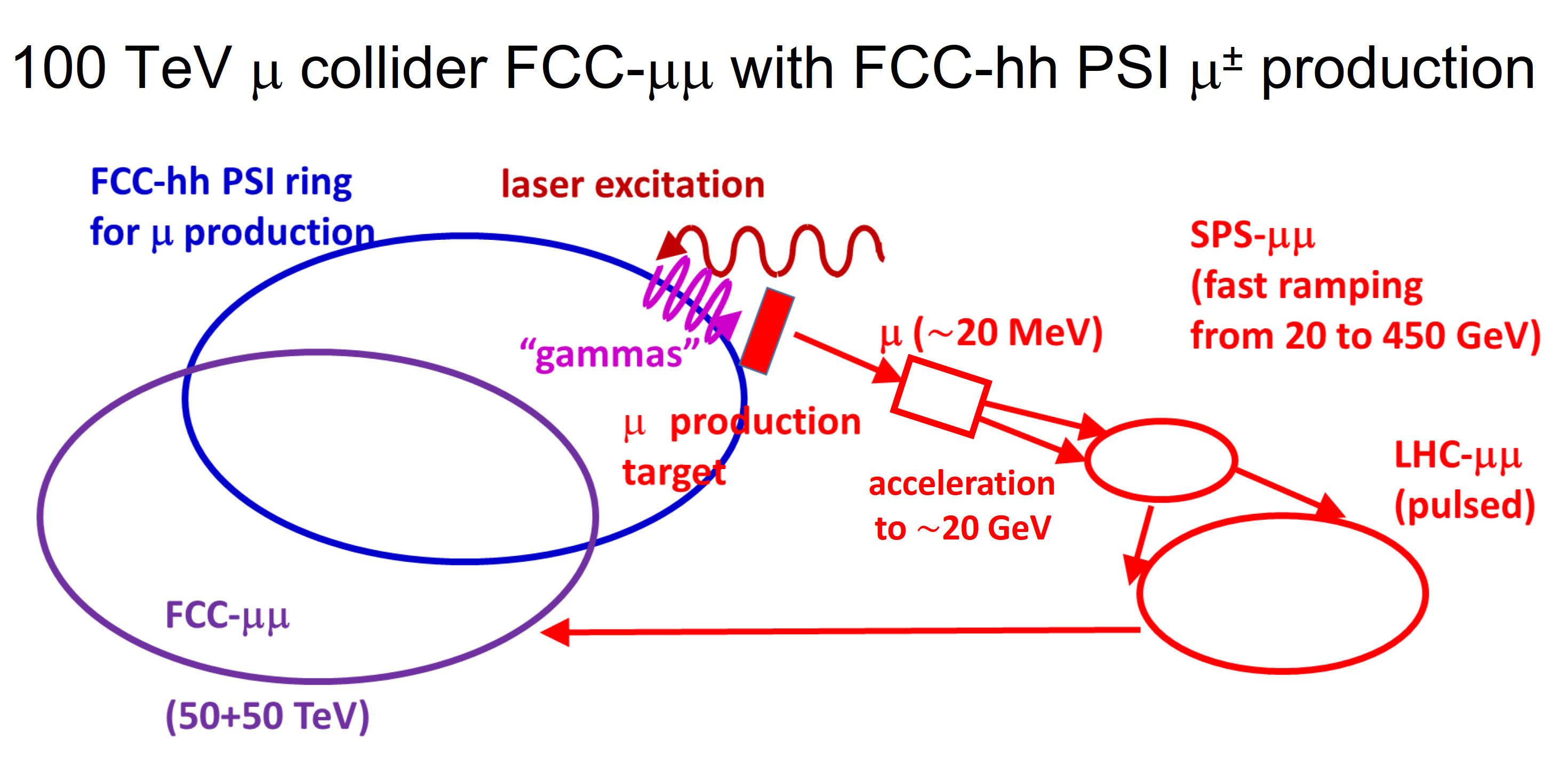}
\includegraphics[width=0.90\columnwidth]{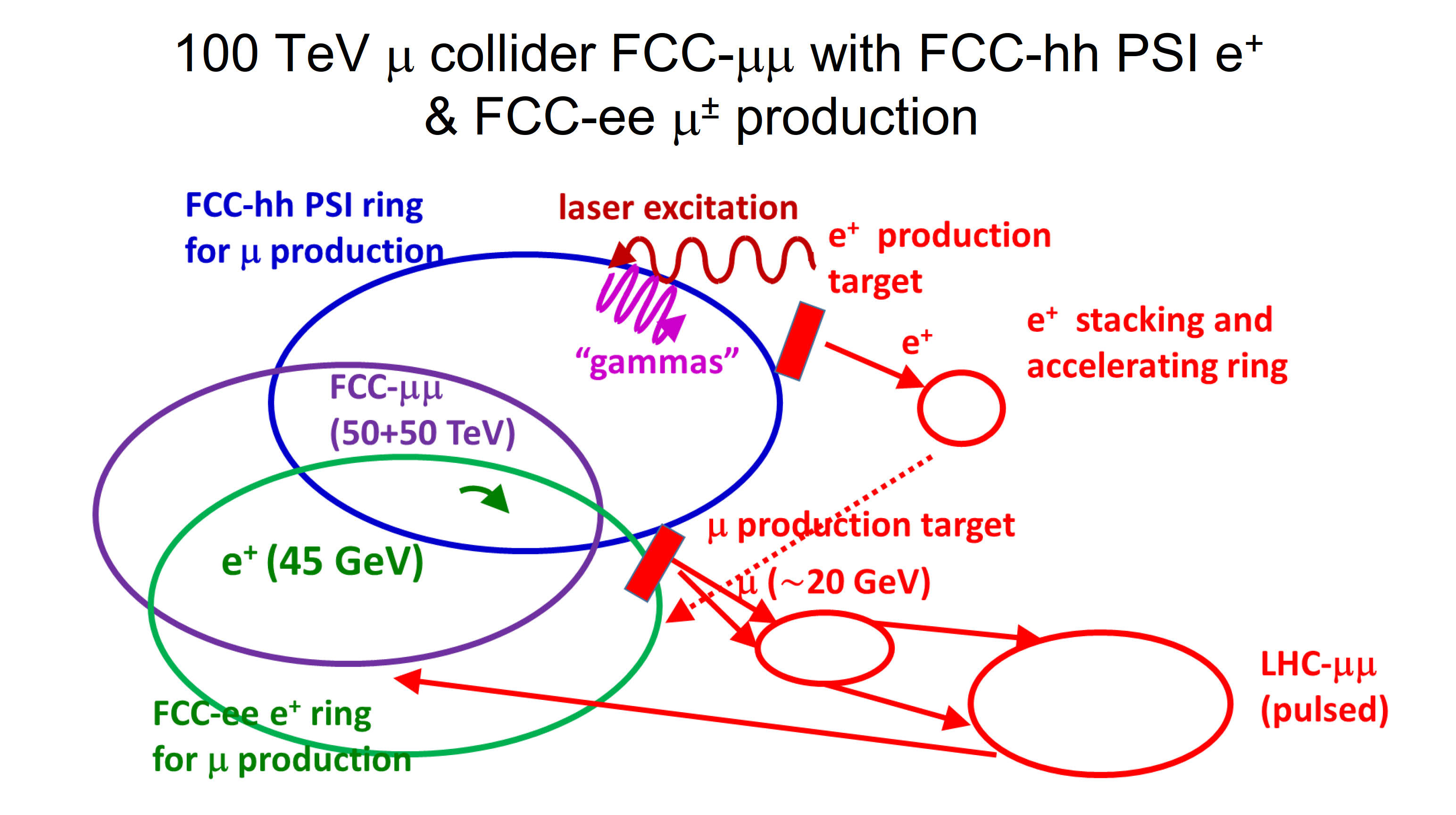}
\caption{Example variants of 
multi-TeV muon colliders based on the LHC/FCC complex.}
\label{fig4}
\end{figure}

Figure \ref{fig4} sketches various example configurations for converting the  
LHC\--FCC complex into a muon collider, using the concepts of low-emittance muon beam generation 
by positron annihilation in a $\sim$45-GeV storage ring \cite{raimondi} 
(which could be FCC-ee or its top-up booster, already optimized for Z pole operation
at the same energy), and/or
of either positron or muon production in laser-Compton 
collisions with a partially stripped heavy-ion
(PSI) beam, the so-called Gamma Factory \cite{krasny}.  
The muon beams would be collided in one of the two LHC or FCC rings,
while the second ring (or the other one of the two accelerators, respectively) 
 might potentially 
be used to store partially stripped heavy-ion beams for muon or positron production.
Remaining issues, still to be tackled, include the muon stacking, 
rapid acceleration, muon synchrotron radiation and neutrino radiation.

\section{Ultimate Limits and Outlook}
An ultimate limit on electromagnetic acceleration may be set by the 
Sauter-Schwinger critical field, above which the QED vacuum breaks down.
This critical field is equal to
$E_{\rm cr}\approx 10^{12}$~MV/m or $B_{\rm ct}\approx 4.4\times 10^{9}$~T.
Assuming these fields, 
the Planck scale of $10^{28}$~eV can be reached 
by a circular or linear collider with a size of about $10^{10}$~m, 
or about a tenth of the distance between earth and sun,
for either type of collider (!). 
A solar-system Planck-energy linear collider was considered previously,  
and it was judged to be  ``not an inconceivable task for an advanced 
technological society'' \cite{chen}.
The presently studied FCC and CEPC/SppC colliders appear small
if compared with a future solar-system collider at the Planck energy.

The future holds other tantalizing accelerator challenges, 
such as  
(1) the possible use of accelerators for the detection or generation of 
gravitational waves \cite{thorne,agnolo,ariesapec}; 
(2) the possibility of constructing a high-field
crystal \cite{shiltsev,scandale} 
or nano-tube \cite{youngmin} accelerator (for acceleration \cite{shiltsev}
or bending \cite{scandale});
(3) the strategy for approaching (or exceeding) 
the Planck scale \cite{beyondplanck}, and
the possible acceleration beyond the Schwinger limit 
e.g., by developing a ``quantum plasma accelerator'' \cite{zimmermann},    
or using techniques of entanglement \cite{brooks}.

\section*{Acknowledgements}
I would like to thank the organizers of EAAC2017,
especially Edda Gschwendtner, Ralph A{\ss}mann, and Massimo Ferrario,  
 for inviting my contribution and proposing the title. 

This work was supported, in part, 
by the European Commission under the HORIZON2020 Integrating Activity project ARIES, grant agreement 730871.

\bibliographystyle{model1-num-names} 
\bibliography{mybib}{}

\end{document}